\documentclass[10pt,aps,prl,twocolumn,showkeys,superscriptaddress]{revtex4-2}

\usepackage{placeins}                   
\usepackage[utf8]{inputenc}
\usepackage{newunicodechar}
\newunicodechar{，}{,}
\usepackage[artemisia]{textgreek}       
\usepackage{amssymb}       
\usepackage{gensymb}
\usepackage{amsmath} 				    
\usepackage{latexsym}                   
\usepackage{nicefrac}                   
\usepackage{float}
\usepackage{graphicx}                   
\usepackage{epsfig}                     
\usepackage{epstopdf}                   
\usepackage{dcolumn}                    

\usepackage{lipsum}                     
\usepackage{color}                      

\usepackage{subfigure}                  
\usepackage{blindtext}                  
\usepackage[ulem=normalem]{changes}     
\usepackage{float}                      
\usepackage{array}                      
\usepackage{soul}                       
\usepackage[T1]{fontenc}                
\usepackage[separate-uncertainty=true,multi-part-units=single]{siunitx}
\usepackage{siunitx}
\usepackage[hidelinks]{hyperref}        

\DeclareSIUnit\sq{\ensuremath{\Box}}
\DeclareSIUnit\bar{bar}
\DeclareSIUnit\angstrom{\text{Å}}
\DeclareSIUnit{\atpercent}{at\%}

\newcolumntype{P}[1]{>{\centering\arraybackslash}p{#1}}
\hypersetup{
    colorlinks,
    linkcolor={red!50!black},
    citecolor={blue!50!black},
    urlcolor={blue!80!black}
}

\newcolumntype{L}[1]{>{\raggedright\arraybackslash}p{#1}}
\newcolumntype{C}[1]{>{\centering\arraybackslash}p{#1}}
\newcolumntype{R}[1]{>{\raggedleft\arraybackslash}p{#1}}

\begin{document}

\title{An InAsSb surface quantum well with in-situ deposited Nb as a platform for semiconductor-superconductor hybrid devices}

\author{Sjoerd Telkamp}
\altaffiliation{These authors contributed equally to this work.}
\affiliation{Solid State Physics Laboratory, ETH Z\"urich, CH-8093 Z\"urich, Switzerland}
\affiliation{Quantum Center, ETH Z\"urich, CH-8093 Z\"urich, Switzerland}
\author{Zijin Lei}
\altaffiliation{These authors contributed equally to this work.}
\affiliation{Solid State Physics Laboratory, ETH Z\"urich, CH-8093 Z\"urich, Switzerland}
\affiliation{Quantum Center, ETH Z\"urich, CH-8093 Z\"urich, Switzerland}

\author{Tommaso Antonelli}
\affiliation{Solid State Physics Laboratory, ETH Z\"urich, CH-8093 Z\"urich, Switzerland}
\affiliation{Quantum Center, ETH Z\"urich, CH-8093 Z\"urich, Switzerland}
\author{Christian Reichl}
\affiliation{Solid State Physics Laboratory, ETH Z\"urich, CH-8093 Z\"urich, Switzerland}
\affiliation{Quantum Center, ETH Z\"urich, CH-8093 Z\"urich, Switzerland}
\author{Ilya Besedin}
\affiliation{Solid State Physics Laboratory, ETH Z\"urich, CH-8093 Z\"urich, Switzerland}
\affiliation{Quantum Center, ETH Z\"urich, CH-8093 Z\"urich, Switzerland}
\affiliation{ETH Zurich—PSI Quantum Computing Hub, Paul Scherrer Institute, CH-5232 Villigen, Switzerland}
\author{Georg Jakobs}
\affiliation{Solid State Physics Laboratory, ETH Z\"urich, CH-8093 Z\"urich, Switzerland}
\affiliation{Quantum Center, ETH Z\"urich, CH-8093 Z\"urich, Switzerland}

\author{Stefan Fält}
\affiliation{Solid State Physics Laboratory, ETH Z\"urich, CH-8093 Z\"urich, Switzerland}
\affiliation{Quantum Center, ETH Z\"urich, CH-8093 Z\"urich, Switzerland}
\author{Christian Marty}
\affiliation{Solid State Physics Laboratory, ETH Z\"urich, CH-8093 Z\"urich, Switzerland}
\affiliation{Quantum Center, ETH Z\"urich, CH-8093 Z\"urich, Switzerland}
\author{Rüdiger Schott}
\affiliation{Solid State Physics Laboratory, ETH Z\"urich, CH-8093 Z\"urich, Switzerland}
\affiliation{Quantum Center, ETH Z\"urich, CH-8093 Z\"urich, Switzerland}

\author{Werner Wegscheider}
\affiliation{Solid State Physics Laboratory, ETH Z\"urich, CH-8093 Z\"urich, Switzerland}
\affiliation{Quantum Center, ETH Z\"urich, CH-8093 Z\"urich, Switzerland}

\begin{abstract}
We present a novel semiconductor-superconductor hybrid material based on a molecular beam epitaxially grown InAsSb surface quantum well with an in-situ deposited Nb top layer. Relative to conventional Al-InAs based systems, the InAsSb surface quantum well offers a lower effective mass and stronger spin-orbit interaction, while the Nb layer has a higher critical temperature and a larger critical magnetic field. The in-situ deposition of the Nb results in a high-quality interface that enables strong coupling to the InAsSb quantum well. Transport measurements on Josephson junctions reveal an induced superconducting gap of 1.3 meV. Furthermore, a planar asymmetric SQUID is realized, exhibiting gate-tunable superimposed oscillations originating from both the individual Josephson junction and the full SQUID loop. The large induced superconducting gap combined with strong spin-orbit interaction position this material as an attractive platform for experiments exploring gate-tunable superconductivity and topological superconducting devices. 
    
\end{abstract}
\maketitle

Semiconductor-superconductor quantum materials provide a versatile platform for realizing gate-tunable superconductor devices and exploring potential topological superconductivity. Devices based on these material systems typically rely on a strong hybridization of superconductors to semiconductors, which have a tunable charge carrier density, large \textit{g}-factors, and strong spin-orbit interactions (SOI). Such properties have enabled a variety of experimental implementations, including Andreev spin qubits \cite{Zazunov2003,Hays2021,Pita-Vidal2023},  superconducting electronic devices, such as Josephson diodes or superconducting rectifiers \cite{Baumgartner2022,Lotfizadeh2024,Ingla_Aynes2025,Kochan2025,Castellani2025}, and investigations into Majorana bound states \cite{vanmourik2012,Albrecht2016,TenHaaf2024,Dvir2023}. 

A widely used approach to achieve these properties is to couple superconductors to two-dimensional electron gases (2DEGs) based on semiconductor quantum wells (QWs), which can provide high mobilities \cite{Shabani2016,Kjaergaard2015,Thomas2019}. Among available material platforms, InAs- and InSb-based QWs are particularly attractive due to their strong and tunable SOI and  large \textit{g}-factors of approximately 15 and 50, respectively. Furthermore, strong coupling to superconductors such as Al \cite{Chang2015,Shabani2016,Cheah2023}, Pb \cite{Kanne2021}, Sn \cite{Pendharkar2021} and Nb \cite{Telkamp2024} has been achieved by realizing high-quality semiconductor-superconductor interfaces. These interfaces enable a large superconducting gap to be induced in the semiconductor. High-quality interfaces are typically realized through in-situ deposition of the superconductor and require a high degree of atomic ordering without intermixing or impurities. 

Recently, the tertiary compound InAs$_{1-x}$Sb$_x$ has emerged as a promising alternative to binary InAs and InSb, since it has been shown to offer even larger \textit{g}-factors and stronger SOI \cite{sarney2017bulk,suchalkin2018engineering,jiang2022giant,Jiang2023}. The incorporation of As into the alloy not only stabilizes the interface but also facilitates strong coupling to superconductors\cite{sarney2020aluminum,Sestoft2018,Mayer2020,Moehle2021}. Notably, both the \textit{g}-factor and SOI are highly sensitive to the mole fraction of As \cite{sarney2017bulk,metti2022}. Furthermore, initial studies have demonstrated transparent interfaces with Al superconducting layers \cite{Mayer2020,Moehle2021}. Although Al enables high-quality epitaxial growth, its relatively small superconducting gap limits device performance because of the associated small critical current, low critical magnetic field, and low critical temperature. To overcome these constraints, alternative superconductors with larger gaps are highly desirable, yet only a few candidates have been explored to date \cite{Khan2023}.

In this work, we present a new semiconductor-superconductor hybrid material platform that combines the large superconducting gap of Nb with the strong SOI of an InAsSb surface QW. A Nb superconducting film is deposited in-situ directly onto the InAsSb QW, enabling the large superconducting gap of Nb to be induced into the semiconductor heterostructure. This approach resulted in a high-quality superconductor-semiconductor interface without significant intermixing. Using this material stack, we fabricate Josephson junctions (JJs) and investigate their low-temperature transport properties. From these measurements, we demonstrate the basic functionality of the device and determine that the induced gap $\Delta$ is 1.34 meV. Finally, we extend our study to a superconducting quantum interference device (SQUID) to explore the current-phase relationships in these devices. These results demonstrate the potential of this material as a platform for experiments exploring gate-tunable superconductivity and topological superconducting devices.

The heterostructures are grown on undoped GaAs substrates by molecular beam epitaxy (MBE) along the (100) direction. After a transition from GaAs to GaSb, a graded buffer structure is grown to overcome the lattice mismatch between the QW and the substrate, allowing dislocations to relax before the QW. A virtual substrate of In$_{0.53}$Al$_{0.47}$Sb is grown and also functions as a bottom barrier. Finally, a 21-nm thick InAsSb layer is grown on top which forms the surface QW. The sample is then transferred through an ultra-high vacuum (UHV) tunnel to a metal DC sputtering chamber with a base pressure lower than $2\times10^{-10}$ mbar. A layer of 50 nm Nb is grown at a pressure of $8\times 10^{-3}$ mbar at a power of 125 W. This is followed with a 10 nm NbTi layer, which prevents the oxidation of the Nb layer during device processing. More details of material growth are presented in the Supplementary Materials.

Figures \ref{f1}(a) and (b) show the interface between the QW and superconductors characterized by high angular dark-field scanning transmission electron microscopy (HAADF). As shown in the schematic diagram in Fig. \ref{f1}.(a), 2DEG resides in the top layer of InAsSb, confined by the bottom barrier and semiconductor-Nb interface. Fig. \ref{f1}.(b) presents a high-resolution image of the interface between Nb and InAsSb, showing crystalline layers with very limited observable intermixing. This intermixing is typically observed due to the formation of NbAs in the case of an Nb-InAs interface \cite{Todt2023,Perla2021,Gusken2017,Telkamp2024}, and we speculate that the introduction of Sb atoms limits this effect. Figure \ref{f1}(b)  suggests an epitaxial relationship between Nb and InAsSb over the studied grain domain size of about 20 nm.

The chemical composition of the semiconductor material stack is investigated by X-ray diffraction. In Fig.1(c), the rocking curve of a symmetric (004) direction scan reveals the composition of the heterostructure. The angle has been converted to the lattice constant in the figure. Peaks corresponding to GaAs substrate, GaSb transition layer,  and In$_{0.53}$Al$_{0.47}$Sb virtual substrate are clearly observed. A peak is pronounced at the lattice constant around $6.4$ \r{A}, indicating that the QW has a composition of InAs$_{0.2}$Sb$_{0.8}$. This As-to-Sb ratio is crucial in determining the semiconductor properties such as the SOI and the \textit{g}-factor. 

The properties of both the InAsSb QW and the superconductor are first individually investigated. An InAsSb surface QW with exactly the same structure but without Nb deposition is grown and fabricated into standard Hall bars. The top gates are deposited on a dielectric layer to tune the electron density in the QW. Using standard AC transport measurement techniques, electron mobility was measured over a carrier density range of $0.5 \times 10^{12}$ to $2 \times 10^{12} \,\text{cm}^{-2}$, with a peak mobility of $9.0 \times 10^3 \,\text{cm}^2/\text{Vs}$ observed at a carrier density of approximately $0.75 \times 10^{12} \,\text{cm}^{-2}$. Furthermore, the SOI in the 2DEG is characterized from the weak anti-localization (WAL). These measurements show a tunable Rashba spin-orbit coefficient $\alpha$ with a maximum of around 85 meV$\si{\angstrom}$, which is significantly higher than for InSb QWs \cite{Lei2022,Lei2023}.

DC magneto-transport measurements are performed to characterize the Nb superconductor in the van der Pauw geometry. These measurements show a critical temperature $T_{\mathrm{c}}$ of $9.5 \pm0.1$ K. This exceeds 9.3 K, which is the maximum $T_{\mathrm{c}}$ value of bulk Nb. The higher $T_{\mathrm{c}}$ value originates from the NbTi capping layer, for which slightly higher $T_{\mathrm{c}}$ have been reported \cite{Bellin1969,Roberts1976}. These results indicate a low impurity concentration in our films. Furthermore, the critical field $B_{c2}$ is measured at various temperatures, from which a $B_{c2} > 3.5$ T at 15 mK can be extrapolated. More details on these measurements can be found in the Supplementary Material \ref{app:super}. 

Next, we investigate the coupling between the Nb superconductor and InAsSb QW by low-temperature transport measurements on a planar JJ. The JJ is defined by electron beam lithography combined with wet and dry etching of the semiconductor and Nb layer, respectively. A metal top gate is deposited on a dielectric layer to tune the electron density between the superconducting leads. More details of device processing are presented in the supplementary material. Fig. \ref{f2} (a) shows a false color SEM image of the device prior to the deposition of the metal top gate. The JJ is approximately 100 nm in length and 2.0 $\mu$m in width and has four leads. With the characterization of the semiconductor part, we calculate that the mean free path of the surface QW reaches $122 \: \rm{nm}$, which is comparable to the length of the JJ. Therefore, the JJ is working in a quasi-ballistic regime. Two leads are used to apply a current through the junction and the other two are used to measure the voltage drop $V_{\rm{DC}}$ across the device. The location of the top gate is illustrated with the blue color in the figure. The transport measurement is performed using standard AC lock-in techniques, where the excitation current $I_{\rm{AC}}$ is 10 nA and the frequency is 83 Hz. All measurements were performed in a dilution refrigerator with a base temperature below 15 mK. 

Figure \ref{f2}(b) shows the differential resistance $R=dV/dI$ plotted as a function of the source-drain current $I_{\rm{SD}}$ and the top gate voltage $V_{\rm{TG}}$. As $V_{\rm{TG}}$ is decreased, the critical current $I_{\rm{C}} $ is reduced, reaching complete pinch-off around $V_{\rm{TG}} = -1.9 \:\rm{V}$. This indicates the electrical tunability of the carrier density between the superconducting leads. Furthermore, this also verifies that there are no parallel current paths in the device. 

We now determine the value of the induced gap in the semiconductor by studying multiple Andreev reflections (MARs). Figure \ref{f2}(c) shows $R$ as a function of $I_{\rm{SD}}$ for various temperatures ranging from 15 mK to 10 K. We note that the switching current at 15 mK is around 7.5 $\mu$A, which results in a critical current density of  $3.75\times10^3 \rm{A}/\rm{cm}^2$ that is more than one order of magnitude larger than conventional Al-based planar JJs \cite{Kjaergaard2015,Cheah2023,Haxell2022}. In Fig. 2 (c), the first three peaks in $R$ that we attribute to MAR are indicated. As the temperature increases, these peaks move towards a smaller $I_{\rm{SD}}$ regime. A supercurrent persists at zero bias up to 7 K, and a clear dip around zero bias persists up to 8 K. This indicates the large energy scales involved in our JJ. 

\begin{figure}
 
    \centering
    \includegraphics[width = 1\linewidth]{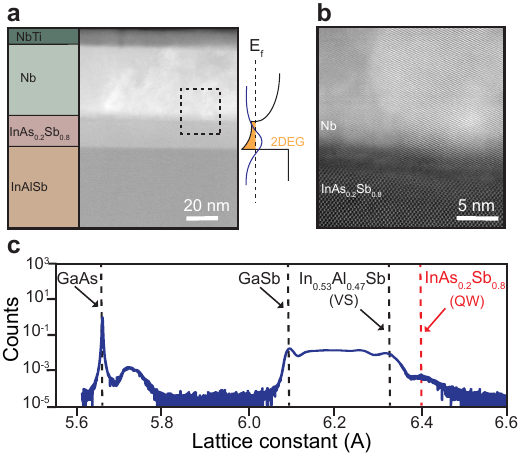}
    
    \caption{
    (a) STEM overview of the top part of the heterostructure,  showing the InAsSb/InAlSb surface QW and the Nb superconducting layers. Schematic sketches of the layers and electron confinement in the surface QW are shown on the left and right of the figure, respectively.  (b) Zoom in of the framed area in (a) with higher resolution. (c) XRD measurements along the (004) direction to determine the components of each layer. The x-axis has been converted from angle to lattice constant.}
    \label{f1}
\end{figure}

\begin{figure*}
\centering
\setlength\fboxsep{0pt}
\setlength\fboxrule{0pt}
\fbox{\includegraphics[width=6.5in]{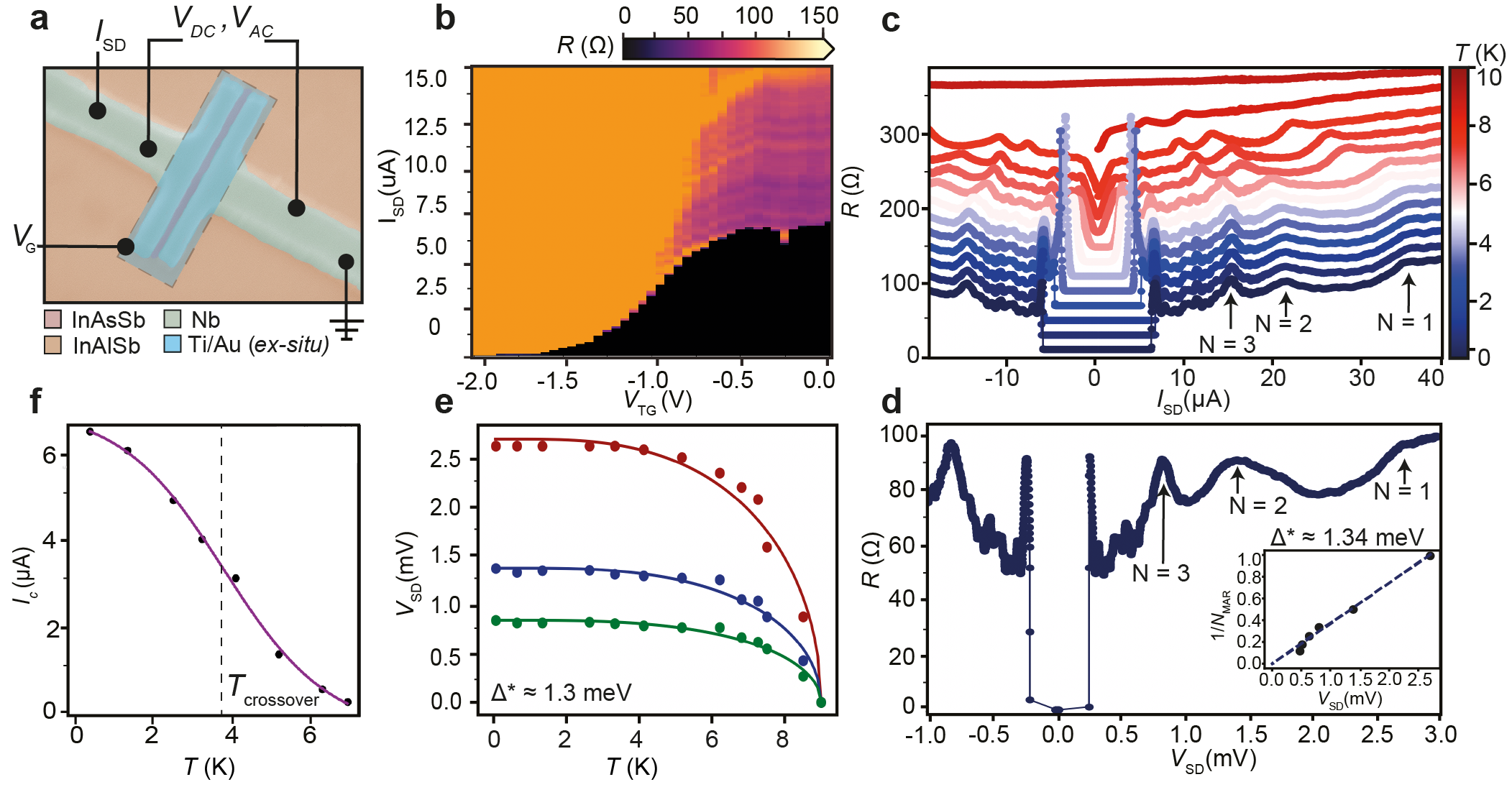}}
\caption{(a) False-color SEM image of the device indicating the source-drain electrodes, top-gate as well as the measured AC voltage and DC voltage. (b) The measured resistance $R$ as a function of  bias current and gate voltage. (c) Temperature dependent measurement of $R-I_{\rm{SD}}$ diagrams. Each trace is offset by 50 $\Omega$ for clarity. The peaks associated to the first three MAR resonances are indicated in the figure.  (d) $R$ as a function of measured $V_{\rm{SD}}$ at 15 mK. Indicated are the MAR peaks. The inset shows the first six MAR peaks plotted as a function of 1/$N_{\rm MAR}$. The dashed line represents the fit associated with an induced gap of 1.34 meV. (e) The temperature dependence of $V_{\rm{SD}}$ where MAR peaks appear. $V_{\rm{SD}}$ for MAR with $N_{\rm MAR}$=1  (red) ,2 (blue), and 3 (green) are plotted (dots) and fitted with equation \ref{BCSstar} (lines). (f) The switching current of the device plotted as a function of temperature. The solid line is the fit of the combined model for the escape mechanism of the JJ. The dashed line indicates the crossover temperature between two regimes of different dominating mechanisms.}
\label{f2}
\end{figure*}

The induced superconducting gap $\Delta ^*$ can be calculated from the voltage at which the MAR characteristics occur. In Fig. \ref{f2}(d), $R$ measured at 15 mK is plotted against $V_{\rm {SD}}$. The inset of the figure shows the values $V_{\rm {SD}}$ at which the first six MAR peaks occur as a function of 1/$N_{\rm MAR}$. This is fitted with the relation $ eV=2 \Delta^* / N_{\rm{MAR}}$,  from which we determine $\Delta ^*$ to be 1.34 meV.  The calculation of $\Delta ^*$ can also be verified by investigating the temperature dependence of the first three MAR peaks. In Fig. \ref{f2}(e),  the $V_{\rm{SD}}$ corresponding to the peaks of $N_{\rm{MAR}}=1,2,$ and 3 are plotted as a function of temperature and the solid lines are fits to the BCS relation \cite{Bardeen1957}
\begin{equation}
\Delta^*(T)=\Delta^*(0) \tanh \left(1.74 \sqrt{\frac{T_{\mathrm{c}}}{T}}-1\right).
\label{BCSstar}
\end{equation}
Using $T_{\rm c}=9$ K obtained from van der Pauw measurements of the same chip and the fitting of $\Delta^*(T)$, we obtain that $\Delta^*$ is approximately 1.3 meV at $T=0$ K. This agrees well with the value previously found for $\Delta^*$.
Lastly, in Fig. \ref{f2}(f) we have plotted the switching current as a function of temperature, enabling a quantitative study of the phase escape mechanism in the JJ. A combined model is used here to include both the macroscopic quantum tunneling (MQT) mechanism at low temperature and the thermally activated (TA) phase diffusion mechanism at high temperature. With the analytical description of the model (details in the Supplementary Material), the crossover temperature describing the transition from MQT to TA is fitted to be $\sim3.8$ K. This means that our junction is well into the quantum regime, even at higher temperatures that could be accessible without a dilution refrigerator.

The induced gap of 1.3 meV close to the BCS-bulk Nb gap of about $K_{\rm{b}}T_{\rm{c}}=1.45$ meV. This implies that the transparency of the semiconductor-superconductor interface is high, but not as good as that of the state-of-the-art Al-based hybrid devices. In Al-based hybrids, almost the full bulk gap can be induced in the semiconductor heterostructure, which is related to the sharp interface achievable in that material platform \cite{Shabani2016,Cheah2023}. This agrees with our characterization of the interface shown in Fig. \ref{f1}. The product of $I_{\rm{c}}R_{\rm{N}}$ can be used to estimate the induced gap and is generally
considered as a figure of merit for the transparency of the barrier \cite{Kjaergaard2015,Shabani2016,DeVries2018}. In this device, the critical current $I_{\rm{c}} =$ 8 $\rm{\mu A}$ and the normal resistance $R_{\rm{N}} = 100 $ $\rm{\Omega}$,  yielding a gap of about 0.8 meV. This indicates a lower interface transparency compared to Al-based devices. However, the magnitude of the superconducting gap is about eight times larger than for conventional hybrid systems \cite{Cheah2023,Kjaergaard2017,Lee2019}. Furthermore, the critical current density, which can reach up to $5.8\times10^3 \rm{A}/\rm{cm}^2$ for a 4 $\mu$m wide JJ (see the supplementary materials \ref{App:otherJJ}), is also about one order of magnitude higher than expected for Al-based devices of similar geometry. These advantages make our material interesting for applications based on gate-tunable supercurrent that would benefit from large critical currents, such as Josephson field effect transistors (joFETs)\cite{Kousar_2022,Vigneau2019,connel2021}, gate-tunable parametric amplifiers \cite{phan2023} and gate-tunable superconducting diodes \cite{ciaccia2023,Caraiola2024,telkamp2025}.

\begin{figure*}
\centering
\setlength\fboxsep{0pt}
\setlength\fboxrule{0pt}
\fbox{\includegraphics[width=6.5in]{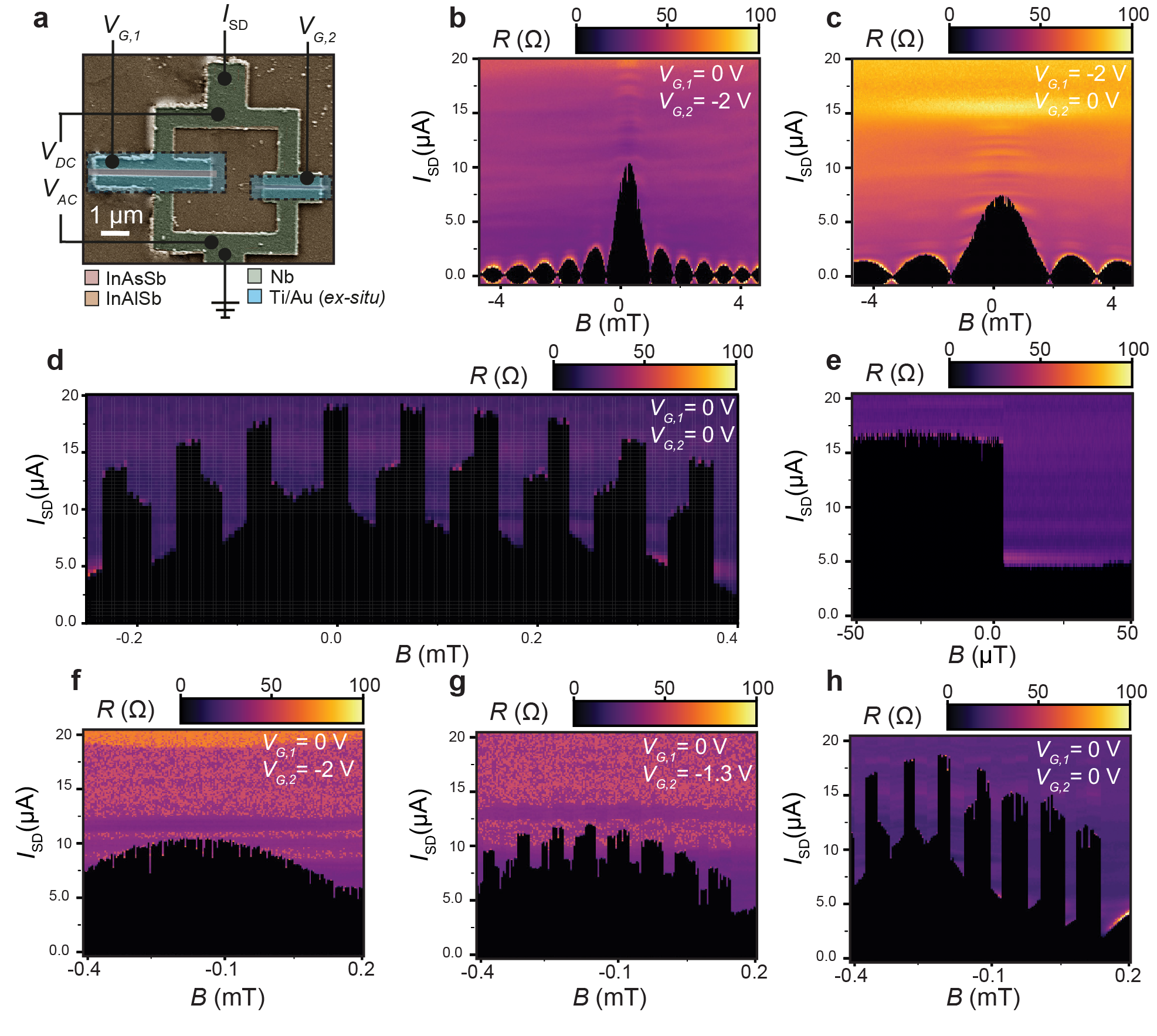}}
\caption{
(a) A false-colored SEM image of the a representative SQUID device on the same chip before gate deposition. The measurement circuit configuration is illustrated. (b) Measurement on the large JJ while the small JJ on the SQUID is pinched off. $R$ is plotted as a function of applied magnetic field $B$ and source-drain bias current $I_{\rm{{SD}}}$. (c) Same as in \textbf{b} but for the small JJ. (d) The interference pattern observed with both junctions fully open. (e) A high resolution measurement of a small area of $B$, indicating the abrupt changes in switching current.  (f)-(h) The interference patterns of the SQUID for a narrow region of magnetic field with the small JJ always open, but the large JJ completely pinched off (f), partially closed (g), and completely open (h).}
\label{f3}
\end{figure*}

To further demonstrate the suitability of our developed material platform for more advanced quantum devices, we show an asymmetric SQUID in Fig. \ref{f3}. Figure \ref{f3}(a) shows a false-colored SEM image of a representative asymmetric SQUID before gate deposition on the same chip as the JJ. The device consists of two JJs, similar to Fig. \ref{f2}, which are 2 $\mu$m by 100 nm and 4 $\mu$m by 200 nm, respectively. The two JJ are connected in a loop with a surface area of 13 $\mu$m$^2$. DC voltages $V_{\rm{G,1}}$ and $V_{\rm{G,2}}$ to the two top gates tune the charge carrier density at each junction separately. The dependence of the supercurrent on the gate voltage for both individual junctions is shown in the supplementary \ref{app:squid}.

Figure \ref{f3}(b) and (c) show $R$ as a function of $I_{\rm{SD}}$ and an out-of-plane magnetic field $B$ when current is only allowed to pass through one of the JJs. Fraunhofer diffraction patterns that are close to ideal appear in both the small (Fig. \ref{f3}(b) and large JJ (Fig. \ref{f3}(c)). The lobes decay as $1/B$ symmetrically in both positive and negative magnetic fields, and the devices show complete destructive interference at the minima. For both junctions, up to the sixth-order maxima can be observed. Based on the JJ geometry and considering the London penetration depth $\lambda_{\rm{L}}=40\: \rm{nm}$ \cite{maxfield1965}, the spacing between the minima should be 1.9 mT and 10 mT for Fig. \ref{f3}(b) and (c) respectively. The observed spacing is smaller, with 0.7 mT and 1.7 mT, indicating a significant contribution of the commonly observed flux-focusing effect \cite{Suominen2017,Cheah2023,Telkamp2024}. 

Fig. \ref{f3}(d) shows the interference pattern measured with both junctions fully open ($V_{G,1} = V_{G,2} =0$). The pattern consists of two superimposed oscillations. One has a larger period of about 0.7 mT, corresponding to the small JJ, and the other has a much smaller period of $\sim80\: \mu$ T, corresponding to threading a single flux quantum through the SQUID loop. The magnitude of the oscillations varies between 7.5 $\mu$ A and 10 $\mu$ A, which is in agreement with the $I_{\rm{c}}$ of the individual JJs. 

In contrast to the Fraunhofer patterns obtained from individual junctions, the oscillations of supercurrent through the full SQUID loop are very abrupt. This is further shown in Fig. \ref{f3}(e), which shows the abrupt step in higher resolution. We speculate that these features relate to switching events between metastable states of the SQUID \cite{goldman1965,MATSINGER197891,seguin1992,thomann2009}. In this perspective, the jumps are enabled by finite loop inductance and large critical currents. They occur when the phase difference across the junctions abruptly changes. As the external magnetic flux is varied, the system minimizes its total energy by discontinuously switching between the metastable states. This results in sharp changes in the supercurrent at intervals of one flux quantum. 

The onset of the behavior is further shown in Figs. \ref{f3}(f),(g) and (h) that offer a direct comparison of the interference pattern when one junction is closed, slightly open or both are fully open, respectively. In Fig. \ref{f3}(f), only a Fraunhofer pattern corresponding to JJ2 is observed, while in Fig. \ref{f3}(h), the interference pattern with a periodicity corresponding to the SQUID dimension is pronounced. Similar deviations from conventional interference patterns are often seen in SQUID devices, especially if the two junctions are asymmetric \cite{Li2024,Zhang2024,perego2024}.

In summary, we have developed a new semiconductor-superconductor hybrid material based on InAsSb surface QW with an in-situ deposited Nb superconducting layer. High-quality interfaces are observed through STEM analysis, with limited intermixing. Low-temperature transport measurements on JJs reveal strong coupling of the surface QW with the superconductor. We observe gate-tunable supercurrent and a near-ideal Fraunhofer pattern. By detailed analysis of the MAR features, the induced gap in the semiconductor is determined to be about 1.3 meV. Furthermore, we highlight key advantages of our JJs, such as its large critical current and the elevated temperature at which superconductivity persists. Finally, a SQUID is realized, which shows two superimposed interference patterns relating magnetic flux through the individual JJ and the full device loop. These results highlight InAsSb 2DEGs with large-gap superconductors as a versatile platform for tunable superconductivity and future topological quantum technologies.


S. T. and Z. L. contributed equally to this work. S. T. performed the superconductor deposition, characterization, and measurements of the hybrid devices. Z. L. was responsible for the semiconductor growth, characterization, and device fabrication. T.A., C.R., S.F. and R.S. supported the growth. G.J. performed the Hall bar measurements and C.M. supported fabrication. W.W. supervised the work. We appreciate fruitful discussions with Prof. T. Ihn. Prof. D. Geshkenbein, Dr. Fabrizio Nichele, Dr. Marco Coraiola and Dr. F. Krizek. We also thank the FIRST-Center for Micro- and Nanoscience at ETH Zurich and Binnig and Rohrer Nanotechnology Center (BRNC) for the support in device fabrication. We acknowledge financial support from the Swiss National Science Foundation (SNSF) and the NCCR QSIT (National Center of Competence in Research - Quantum Science and Technology).
\newpage

\bibliographystyle{ieeetr}
\bibliography{papers}

\begin{thebibliography}{10}

\bibitem{Zazunov2003}
A.~Zazunov, V.~S. Shumeiko, E.~N. Bratus', J.~Lantz, and G.~Wendin, ``Andreev level qubit,'' {\em Phys. Rev. Lett.}, vol.~90, p.~087003, Feb 2003.

\bibitem{Hays2021}
M.~Hays, V.~Fatemi, D.~Bouman, J.~Cerrillo, S.~Diamond, K.~Serniak, T.~Connolly, P.~Krogstrup, J.~Nyg{\aa}rd, A.~{Levy Yeyati}, A.~Geresdi, and M.~H. Devoret, ``{Coherent manipulation of an Andreev spin qubit},'' {\em Science}, vol.~373, no.~6553, pp.~430--433, 2021.

\bibitem{Pita-Vidal2023}
M.~Pita-Vidal, A.~Bargerbos, R.~{\v{Z}}itko, L.~J. Splitthoff, L.~Gr{\"{u}}nhaupt, J.~J. Wesdorp, Y.~Liu, L.~P. Kouwenhoven, R.~Aguado, B.~van Heck, A.~Kou, and C.~K. Andersen, ``{Direct manipulation of a superconducting spin qubit strongly coupled to a transmon qubit},'' {\em Nat. Phys.}, vol.~19, no.~8, pp.~1110--1115, 2023.

\bibitem{Baumgartner2022}
C.~Baumgartner, L.~Fuchs, A.~Costa, S.~Reinhardt, S.~Gronin, G.~C. Gardner, T.~Lindemann, M.~J. Manfra, P.~E. {Faria Junior}, D.~Kochan, J.~Fabian, N.~Paradiso, and C.~Strunk, ``{Supercurrent rectification and magnetochiral effects in symmetric Josephson junctions},'' {\em Nat. Nanotech.}, vol.~17, no.~1, pp.~39--44, 2022.

\bibitem{Lotfizadeh2024}
N.~Lotfizadeh, W.~F. Schiela, B.~Pekerten, P.~Yu, B.~H. Elfeky, W.~M. Strickland, A.~Matos-Abiague, and J.~Shabani, ``{Superconducting diode effect sign change in epitaxial Al-InAs Josephson junctions},'' {\em Comm. Phys.}, vol.~7, no.~1, p.~120, 2024.

\bibitem{Ingla_Aynes2025}
J.~Ingla-Ayn{\'{e}}s, Y.~Hou, S.~Wang, E.-D. Chu, O.~A. Mukhanov, P.~Wei, and J.~S. Moodera, ``{Efficient superconducting diodes and rectifiers for quantum circuitry},'' {\em Nat. Electron.}, vol.~8, no.~5, pp.~411--416, 2025.

\bibitem{Kochan2025}
D.~Kochan and C.~Strunk, ``{Low-loss electronics with superconducting diodes},'' {\em Nat. Electron.}, vol.~8, no.~5, pp.~380--381, 2025.

\bibitem{Castellani2025}
M.~Castellani, O.~Medeiros, A.~Buzzi, R.~A. Foster, M.~Colangelo, and K.~K. Berggren, ``{A superconducting full-wave bridge rectifier},'' {\em Nat. Electron.}, vol.~8, no.~5, pp.~417--425, 2025.

\bibitem{vanmourik2012}
V.~Mourik, K.~Zuo, S.~M. Frolov, S.~R. Plissard, E.~P. A.~M. Bakkers, and L.~P. Kouwenhoven, ``{Signatures of Majorana Fermions in Hybrid Superconductor-Semiconductor Nanowire Devices},'' {\em Science}, vol.~336, no.~6084, pp.~1003--1007, 2012.

\bibitem{Albrecht2016}
S.~M. Albrecht, A.~P. Higginbotham, M.~Madsen, F.~Kuemmeth, T.~S. Jespersen, J.~Nyg{\aa}rd, P.~Krogstrup, and C.~M. Marcus, ``{Exponential protection of zero modes in Majorana islands},'' {\em Nature}, vol.~531, no.~7593, pp.~206--209, 2016.

\bibitem{TenHaaf2024}
S.~L.~D. ten Haaf, Q.~Wang, A.~M. Bozkurt, C.-X. Liu, I.~Kulesh, P.~Kim, D.~Xiao, C.~Thomas, M.~J. Manfra, T.~Dvir, M.~Wimmer, and S.~Goswami, ``{A two-site Kitaev chain in a two-dimensional electron gas},'' {\em Nature}, vol.~630, no.~8016, pp.~329--334, 2024.

\bibitem{Dvir2023}
T.~Dvir, G.~Wang, N.~van Loo, C.-X. Liu, G.~P. Mazur, A.~Bordin, S.~L.~D. ten Haaf, J.-Y. Wang, D.~van Driel, F.~Zatelli, X.~Li, F.~K. Malinowski, S.~Gazibegovic, G.~Badawy, E.~P. A.~M. Bakkers, M.~Wimmer, and L.~P. Kouwenhoven, ``{Realization of a minimal Kitaev chain in coupled quantum dots},'' {\em Nature}, vol.~614, no.~7948, pp.~445--450, 2023.

\bibitem{Shabani2016}
J.~Shabani, M.~Kjaergaard, H.~J. Suominen, Y.~Kim, F.~Nichele, K.~Pakrouski, T.~Stankevic, R.~M. Lutchyn, P.~Krogstrup, R.~Feidenhans'L, S.~Kraemer, C.~Nayak, M.~Troyer, C.~M. Marcus, and C.~J. Palmstr{\o}m, ``{Two-dimensional epitaxial superconductor-semiconductor heterostructures: A platform for topological superconducting networks},'' {\em Phys. Rev. B}, vol.~93, no.~15, pp.~1--6, 2016.

\bibitem{Kjaergaard2015}
M.~Kjaergaard, ``{Proximity Induced Superconducting Properties in One and Two Dimensional Semiconductors},'' {\em PhD Thesis}, 2015.

\bibitem{Thomas2019}
C.~Thomas, R.~E. Diaz, J.~H. Dycus, M.~E. Salmon, R.~E. Daniel, T.~Wang, G.~C. Gardner, and M.~J. Manfra, ``Toward durable al-insb hybrid heterostructures via epitaxy of 2ml interfacial inas screening layers,'' {\em Phys. Rev. Mater.}, vol.~3, p.~124202, Dec 2019.

\bibitem{Chang2015}
W.~Chang, S.~M. Albrecht, T.~S. Jespersen, F.~Kuemmeth, P.~Krogstrup, J.~Nyg{\aa}rd, and C.~M. Marcus, ``{Hard gap in epitaxial semiconductor-superconductor nanowires},'' {\em Nat. Nanotechnol.}, vol.~10, no.~3, pp.~232--236, 2015.

\bibitem{Cheah2023}
E.~Cheah, D.~Z. Haxell, R.~Schott, P.~Zeng, E.~Paysen, S.~C. ten Kate, M.~Coraiola, M.~Landstetter, A.~B. Zadeh, A.~Trampert, M.~Sousa, H.~Riel, F.~Nichele, W.~Wegscheider, and F.~Krizek, ``Control over epitaxy and the role of the inas/al interface in hybrid two-dimensional electron gas systems,'' {\em Phys. Rev. Mater.}, vol.~7, p.~073403, Jul 2023.

\bibitem{Kanne2021}
T.~Kanne, M.~Marnauza, D.~Olsteins, D.~J. Carrad, J.~E. Sestoft, J.~de~Bruijckere, L.~Zeng, E.~Johnson, E.~Olsson, K.~Grove-Rasmussen, and J.~Nyg{\aa}rd, ``{Epitaxial Pb on InAs nanowires for quantum devices},'' {\em Nat. Nanotechnol. 2021 16:7}, vol.~16, no.~7, pp.~776--781, 2021.

\bibitem{Pendharkar2021}
M.~Pendharkar, B.~Zhang, H.~Wu, A.~Zarassi, P.~Zhang, C.~P. Dempsey, J.~S. Lee, S.~D. Harrington, G.~Badawy, S.~Gazibegovic, R.~L. {Op het Veld}, M.~Rossi, J.~Jung, A.~H. Chen, M.~A. Verheijen, M.~Hocevar, E.~P. Bakkers, C.~J. Palmstr{\o}m, and S.~M. Frolov, ``{Parity-preserving and magnetic field–resilient superconductivity in InSb nanowires with Sn shells},'' {\em Science}, vol.~372, no.~6541, pp.~508--511, 2021.

\bibitem{Telkamp2024}
S.~Telkamp, T.~Antonelli, C.~Todt, M.~Hinderling, M.~Coraiola, D.~Haxell, S.~C. ten Kate, D.~Sabonis, P.~Zeng, R.~Schott, E.~Cheah, C.~Reichl, F.~Nichele, F.~Krizek, and W.~Wegscheider, ``{Development of a Nb-Based Semiconductor-Superconductor Hybrid 2DEG Platform},'' {\em Adv. Electron. Mater.}, vol.~11, no.~7, p.~2400687, 2025.

\bibitem{sarney2017bulk}
W.~Sarney, S.~Svensson, Y.~Xu, D.~Donetsky, and G.~Belenky, ``Bulk {InAsSb} with 0.1 ev bandgap on gaas,'' {\em Journal of Applied Physics}, vol.~122, no.~2, 2017.

\bibitem{suchalkin2018engineering}
S.~Suchalkin, G.~Belenky, M.~Ermolaev, S.~Moon, Y.~Jiang, D.~Graf, D.~Smirnov, B.~Laikhtman, L.~Shterengas, G.~Kipshidze, {\em et~al.}, ``Engineering dirac materials: Metamorphic {InAs1-x Sbx/InAs1-ySby} superlattices with ultralow bandgap,'' {\em Nano Letters}, vol.~18, no.~1, pp.~412--417, 2018.

\bibitem{jiang2022giant}
Y.~Jiang, M.~Ermolaev, G.~Kipshidze, S.~Moon, M.~Ozerov, D.~Smirnov, Z.~Jiang, and S.~Suchalkin, ``Giant g-factors and fully spin-polarized states in metamorphic short-period {InAsSb/InSb} superlattices,'' {\em Nature Communications}, vol.~13, no.~1, p.~5960, 2022.

\bibitem{Jiang2023}
Y.~Jiang, M.~Ermolaev, S.~Moon, G.~Kipshidze, G.~Belenky, S.~Svensson, M.~Ozerov, D.~Smirnov, Z.~Jiang, and S.~Suchalkin, ``$g$-factor engineering with {InAsSb} alloys toward zero band gap limit,'' {\em Phys. Rev. B}, vol.~108, p.~L121201, Sep 2023.

\bibitem{sarney2020aluminum}
W.~L. Sarney, S.~P. Svensson, A.~C. Leff, W.~F. Schiela, J.~O. Yuan, M.~C. Dartiailh, W.~Mayer, K.~S. Wickramasinghe, and J.~Shabani, ``Aluminum metallization of {III--V} semiconductors for the study of proximity superconductivity,'' {\em Journal of Vacuum Science \& Technology B}, vol.~38, no.~3, 2020.

\bibitem{Sestoft2018}
J.~E. Sestoft, T.~Kanne, A.~N. Gejl, M.~von Soosten, J.~S. Yodh, D.~Sherman, B.~Tarasinski, M.~Wimmer, E.~Johnson, M.~Deng, J.~Nyg\aa{}rd, T.~S. Jespersen, C.~M. Marcus, and P.~Krogstrup, ``Engineering hybrid epitaxial inassb/al nanowires for stronger topological protection,'' {\em Phys. Rev. Mater.}, vol.~2, p.~044202, Apr 2018.

\bibitem{Mayer2020}
W.~Mayer, W.~F. Schiela, J.~Yuan, M.~Hatefipour, W.~L. Sarney, S.~P. Svensson, A.~C. Leff, T.~Campos, K.~S. Wickramasinghe, M.~C. Dartiailh, I.~{\v{Z}}uti{\'{c}}, and J.~Shabani, ``{Superconducting Proximity Effect in InAsSb Surface Quantum Wells with In Situ Al Contacts},'' {\em ACS Appl. Electron. Mater.}, vol.~2, pp.~2351--2356, aug 2020.

\bibitem{Moehle2021}
C.~M. Moehle, C.~T. Ke, Q.~Wang, C.~Thomas, D.~Xiao, S.~Karwal, M.~Lodari, V.~{Van De Kerkhof}, R.~Termaat, G.~C. Gardner, G.~Scappucci, M.~J. Manfra, and S.~Goswami, ``{InSbAs two-dimensional electron gases as a platform for topological superconductivity},'' {\em Nano Lett.}, 2021.

\bibitem{metti2022}
S.~Metti, C.~Thomas, D.~Xiao, and M.~J. Manfra, ``Spin-orbit coupling and electron scattering in high-quality ${\mathrm{insb}}_{1\ensuremath{-}x}{\mathrm{as}}_{x}$ quantum wells,'' {\em Phys. Rev. B}, vol.~106, p.~165304, Oct 2022.

\bibitem{Khan2023}
S.~A. Khan, S.~Mart{\'{i}}-S{\'{a}}nchez, D.~Olsteins, C.~Lampadaris, D.~J. Carrad, Y.~Liu, J.~Qui{\~{n}}ones, M.~{Chiara Spadaro}, T.~{Sand Jespersen}, P.~Krogstrup, and J.~Arbiol, ``{Epitaxially Driven Phase Selectivity of Sn in Hybrid Quantum Nanowires},'' {\em ACS Nano}, vol.~17, pp.~11794--11804, jun 2023.

\bibitem{Todt2023}
C.~Todt, S.~Telkamp, F.~Krizek, C.~Reichl, M.~Gabureac, R.~Schott, E.~Cheah, P.~Zeng, T.~Weber, A.~M{\"{u}}ller, C.~Vockenhuber, M.~B. Panah, and W.~Wegscheider, ``{Development of Nb-GaAs based superconductor-semiconductor hybrid platform by combining in situ dc magnetron sputtering and molecular beam epitaxy},'' {\em Phys. Rev. Mater.}, vol.~7, no.~7, p.~76201, 2023.

\bibitem{Perla2021}
P.~Perla, H.~A. Fonseka, P.~Zellekens, R.~Deacon, Y.~Han, J.~K{\"{o}}lzer, T.~M{\"{o}}rstedt, B.~Bennemann, A.~Espiari, K.~Ishibashi, D.~Gr{\"{u}}tzmacher, A.~M. Sanchez, M.~I. Lepsa, and T.~Sch{\"{a}}pers, ``{Fully in situ Nb/InAs-nanowire Josephson junctions by selective-area growth and shadow evaporation},'' {\em Nanoscale Adv.}, vol.~3, no.~5, pp.~1413--1421, 2021.

\bibitem{Gusken2017}
N.~A. G{\"{u}}sken, T.~Rieger, P.~Zellekens, B.~Bennemann, E.~Neumann, M.~I. Lepsa, T.~Sch{\"{a}}pers, and D.~Gr{\"{u}}tzmacher, ``{MBE growth of Al/InAs and Nb/InAs superconducting hybrid nanowire structures},'' {\em Nanoscale Adv.}, vol.~9, no.~43, pp.~16735--16741, 2017.

\bibitem{Lei2022}
Z.~Lei, E.~Cheah, K.~Rubi, M.~E. Bal, C.~Adam, R.~Schott, U.~Zeitler, W.~Wegscheider, T.~Ihn, and K.~Ensslin, ``High-quality two-dimensional electron gas in undoped insb quantum wells,'' {\em Phys. Rev. Res.}, vol.~4, p.~013039, Jan 2022.

\bibitem{Lei2023}
Z.~Lei, E.~Cheah, F.~Krizek, R.~Schott, T.~B\"ahler, P.~M\"arki, W.~Wegscheider, M.~Shayegan, T.~Ihn, and K.~Ensslin, ``Gate-defined two-dimensional hole and electron systems in an undoped insb quantum well,'' {\em Phys. Rev. Res.}, vol.~5, p.~013117, Feb 2023.

\bibitem{Bellin1969}
P.~H. Bellin, V.~Sadagopan, and H.~C. Gatos, ``{Ternary Superconducting Alloys of the Titanium‐Niobium‐Vanadium System. Transition Temperature Variation},'' {\em J. Appl. Phys.}, vol.~40, pp.~3982--3984, sep 1969.

\bibitem{Roberts1976}
B.~W. Roberts, ``{Survey of superconductive materials and critical evaluation of selected properties},'' {\em J. Phys. Chem. Ref. Data}, vol.~5, no.~3, pp.~581--822, 1976.

\bibitem{Haxell2022}
D.~Z. Haxell, E.~Cheah, F.~K\ifmmode \check{r}\else \v{r}\fi{}\'{\i}\ifmmode~\check{z}\else \v{z}\fi{}ek, R.~Schott, M.~F. Ritter, M.~Hinderling, W.~Belzig, C.~Bruder, W.~Wegscheider, H.~Riel, and F.~Nichele, ``Measurements of phase dynamics in planar josephson junctions and squids,'' {\em Phys. Rev. Lett.}, vol.~130, p.~087002, Feb 2023.

\bibitem{Bardeen1957}
J.~Bardeen, L.~N. Cooper, and J.~R. Schrieffer, ``Theory of superconductivity,'' {\em Phys. Rev.}, vol.~108, pp.~1175--1204, Dec 1957.

\bibitem{DeVries2018}
F.~K. de~Vries, J.~Shen, R.~J. Skolasinski, M.~P. Nowak, D.~Varjas, L.~Wang, M.~Wimmer, J.~Ridderbos, F.~A. Zwanenburg, A.~Li, S.~Koelling, M.~A. Verheijen, E.~P. A.~M. Bakkers, and L.~P. Kouwenhoven, ``{Spin–Orbit Interaction and Induced Superconductivity in a One-Dimensional Hole Gas},'' {\em Nano Lett.}, vol.~18, pp.~6483--6488, oct 2018.

\bibitem{Kjaergaard2017}
M.~Kjaergaard, H.~J. Suominen, M.~P. Nowak, A.~R. Akhmerov, J.~Shabani, C.~J. Palmstr\o{}m, F.~Nichele, and C.~M. Marcus, ``Transparent semiconductor-superconductor interface and induced gap in an epitaxial heterostructure josephson junction,'' {\em Phys. Rev. Appl.}, vol.~7, p.~034029, 2017.

\bibitem{Lee2019}
J.~S. Lee, B.~Shojaei, M.~Pendharkar, A.~P. Mcfadden, Y.~Kim, H.~J. Suominen, M.~Kjaergaard, F.~Nichele, H.~Zhang, C.~M. Marcus, and C.~J. Palmstr{\o}m, ``{Transport Studies of Epi-Al/InAs Two-Dimensional Electron Gas Systems for Required Building-Blocks in Topological Superconductor Networks},'' {\em Nano Lett.}, vol.~19, 2019.

\bibitem{Kousar_2022}
B.~Kousar, D.~J. Carrad, L.~Stampfer, P.~Krogstrup, J.~Nygård, and T.~S. Jespersen, ``Inas/more hybrid semiconductor/superconductor nanowire devices,'' {\em Nano Lett.}, vol.~22, p.~8845–8851, Nov. 2022.

\bibitem{Vigneau2019}
F.~Vigneau, R.~Mizokuchi, D.~C. Zanuz, X.~Huang, S.~Tan, R.~Maurand, S.~Frolov, A.~Sammak, G.~Scappucci, F.~Lefloch, and S.~{De Franceschi}, ``{Germanium Quantum-Well Josephson Field-Effect Transistors and Interferometers},'' {\em Nano Lett.}, vol.~19, pp.~1023--1027, feb 2019.

\bibitem{connel2021}
J.~{O'Connell Yuan}, K.~S. Wickramasinghe, W.~M. Strickland, M.~C. Dartiailh, K.~Sardashti, M.~Hatefipour, and J.~Shabani, ``{Epitaxial superconductor-semiconductor two-dimensional systems for superconducting quantum circuits},'' {\em J. Vac. Sci. Technol.}, vol.~39, no.~3, p.~33407, 2021.

\bibitem{phan2023}
D.~Phan, P.~Falthansl-Scheinecker, U.~Mishra, W.~M. Strickland, D.~Langone, J.~Shabani, and A.~P. Higginbotham, ``{Gate-Tunable Superconductor-Semiconductor Parametric Amplifier},'' {\em Phys. Rev. Appl.}, vol.~19, p.~64032, jun 2023.

\bibitem{ciaccia2023}
C.~Ciaccia, R.~Haller, A.~C.~C. Drachmann, T.~Lindemann, M.~J. Manfra, C.~Schrade, and C.~Sch\"onenberger, ``Gate-tunable josephson diode in proximitized inas supercurrent interferometers,'' {\em Phys. Rev. Res.}, vol.~5, p.~033131, Aug 2023.

\bibitem{Caraiola2024}
M.~Coraiola, A.~E. Svetogorov, D.~Z. Haxell, D.~Sabonis, M.~Hinderling, S.~C. ten Kate, E.~Cheah, F.~Krizek, R.~Schott, W.~Wegscheider, J.~C. Cuevas, W.~Belzig, and F.~Nichele, ``{Flux-Tunable Josephson Diode Effect in a Hybrid Four-Terminal Josephson Junction},'' {\em ACS Nano}, vol.~18, no.~12, pp.~9221--9231, 2024.

\bibitem{telkamp2025}
S.~Telkamp, J.~Zhao, and S.~Vaitiekėnas, ``Voltage-tunable field-free josephson diode,'' {\em arXiv:2508.12056}, 2025.

\bibitem{maxfield1965}
B.~W. Maxfield and W.~L. McLean, ``Superconducting penetration depth of niobium,'' {\em Phys. Rev.}, vol.~139, pp.~A1515--A1522, Aug 1965.

\bibitem{Suominen2017}
H.~J. Suominen, J.~Danon, M.~Kjaergaard, K.~Flensberg, J.~Shabani, C.~J. Palmstr\o{}m, F.~Nichele, and C.~M. Marcus, ``Anomalous fraunhofer interference in epitaxial superconductor-semiconductor josephson junctions,'' {\em Phys. Rev. B}, vol.~95, p.~035307, 2017.

\bibitem{goldman1965}
A.~M. Goldman, P.~J. Kreisman, and D.~J. Scalapino, ``Metastable current-carrying states of weakly coupled superconductors,'' {\em Phys. Rev. Lett.}, vol.~15, pp.~495--499, Sep 1965.

\bibitem{MATSINGER197891}
A.~A.~J. Matsinger, R.~{de Bruyn Ouboter}, and H.~van Beelen, ``{On the hysteretic behaviour of a superconducting ring, containing two weak links in an applied magnetic field},'' {\em Physica B C}, vol.~94, no.~1, pp.~91--96, 1978.

\bibitem{seguin1992}
V.~Lefevre-Seguin, E.~Turlot, C.~Urbina, D.~Esteve, and M.~H. Devoret, ``Thermal activation of a hysteretic dc superconducting quantum interference device from its different zero-voltage states,'' {\em Phys. Rev. B}, vol.~46, pp.~5507--5522, Sep 1992.

\bibitem{thomann2009}
A.~U. Thomann, V.~B. Geshkenbein, and G.~Blatter, ``Quantum instability in a dc squid with strongly asymmetric dynamical parameters,'' {\em Phys. Rev. B}, vol.~79, p.~184515, May 2009.

\bibitem{Li2024}
R.~Li, W.~Song, W.~Miao, Z.~Yu, Z.~Wang, S.~Yang, Y.~Gao, Y.~Wang, F.~Chen, Z.~Geng, L.~Yang, J.~Xu, X.~Feng, T.~Wang, Y.~Zang, L.~Li, R.~Shang, Q.~Xue, K.~He, and H.~Zhang, ``{Selective-Area-Grown PbTe-Pb Planar Josephson Junctions for Quantum Devices},'' {\em Nan. Lett.}, vol.~24, pp.~4658--4664, apr 2024.

\bibitem{Zhang2024}
P.~Zhang, A.~Zarassi, L.~Jarjat, V.~V. de~Sande, M.~Pendharkar, J.~S. Lee, C.~P. Dempsey, A.~P. McFadden, S.~D. Harrington, J.~T. Dong, H.~Wu, A.~H. Chen, M.~Hocevar, C.~J. Palmstrøm, and S.~M. Frolov, ``{Large second-order Josephson effect in planar superconductor-semiconductor junctions},'' {\em SciPost Phys.}, vol.~16, p.~030, 2024.

\bibitem{perego2024}
M.~Perego, C.~G. Agero, A.~M. Torà, E.~Portolés, A.~O. Denisov, T.~Taniguchi, K.~Watanabe, F.~Gaggioli, V.~Geshkenbein, G.~Blatter, T.~Ihn, and K.~Ensslin, ``Experimental detection of vortices in magic-angle graphene,'' {\em arXiv:2410.03508}, 2024.

\bibitem{Lehner2018}
C.~A. Lehner, T.~Tschirky, T.~Ihn, W.~Dietsche, J.~Keller, S.~F\"alt, and W.~Wegscheider, ``Limiting scattering processes in high-mobility insb quantum wells grown on gasb buffer systems,'' {\em Phys. Rev. Mater.}, vol.~2, p.~054601, May 2018.

\bibitem{Werthammer1966}
N.~R. Werthamer, E.~Helfand, and P.~C. Hohenberg, ``Temperature and purity dependence of the superconducting critical field, ${H}_{c2}$. iii. electron spin and spin-orbit effects,'' {\em Phys. Rev.}, vol.~147, pp.~295--302, Jul 1966.

\bibitem{Bal2024}
M.~E. Bal, E.~Cheah, Z.~Lei, R.~Schott, C.~A. Lehner, H.~Engelkamp, W.~Wegscheider, and U.~Zeitler, ``Quantum hall effect in inassb quantum wells at elevated temperatures,'' {\em Phys. Rev. Res.}, vol.~6, p.~023259, Jun 2024.

\end{thebibliography}

\appendix

\section{Material growth}
The molecular beam epitaxy (MBE) growth of the InAsSb quantum wells (QWs) follows a procedure similar to that reported in Ref. \cite{Mayer2020}. The process begins with the growth of GaAs and GaAs/Al$_{0.4}$Ga$_{0.6}$As superlattice on an undoped GaAs substrate. A low-temperature transition to GaSb is then carried out, followed by the growth of a GaSb/AlGaSb superlattice. Subsequently, a 3.6 $\rm{\mu}$m-thick graded buffer layer is grown, in which the composition is gradually tuned from In$_{0.02}$Al$_{0.4}$Ga$_{0.58}$Sb to In$_{0.58}$Al$_{0.4}$Ga$_{0.02}$Sb. On top of this, a 300 nm-thick In$_{0.53}$Al$_{0.47}$Sb layer is deposited, serving as both the virtual substrate and the bottom barrier. Finally, a 23-nm thick InAsSb layer is grown to form the surface QW.

In addition to the graded buffer, the lattice mismatch between GaSb and the In$_{0.57}$Al$_{0.43}$Sb virtual substrate can also be accommodated by introducing an In$_{0.4}$Al$_{0.6}$Sb/In$_{0.57}$Al$_{0.43}$Sb superlattice. Further details on the growth conditions are provided in Ref. \cite{Lehner2018}. No significant differences in carrier density or mobility were observed between surface QWs grown using these two different buffer systems.

The Nb films are sputtered using DC magnetron sputtering in a custom-made UHV magnetron sputtering chamber connected to the MBE via a UHV-tunnel \cite{Todt2023}. Samples grown in the MBE are transferred to the magnetron sputtering system and the Nb film is deposited at room temperature using a two-inch target. The films are deposited with a rate of 2.6 \AA/s and at an Ar pressure of $8.8\times 10^{-3}$ mbar during the deposition. The power during the deposition was kept constant at 125 W. A 10 nm thick NbTi capping layer was deposited directly on top of the Nb layer to prevent oxidation during fabrication. A target with a weight ratio of Nb$_{0.68}$Ti$_{0.32}$ was used for the NbTi deposition.

\section{XRD measurement}

\begin{figure}
    \centering
    \includegraphics[width = 1\linewidth]{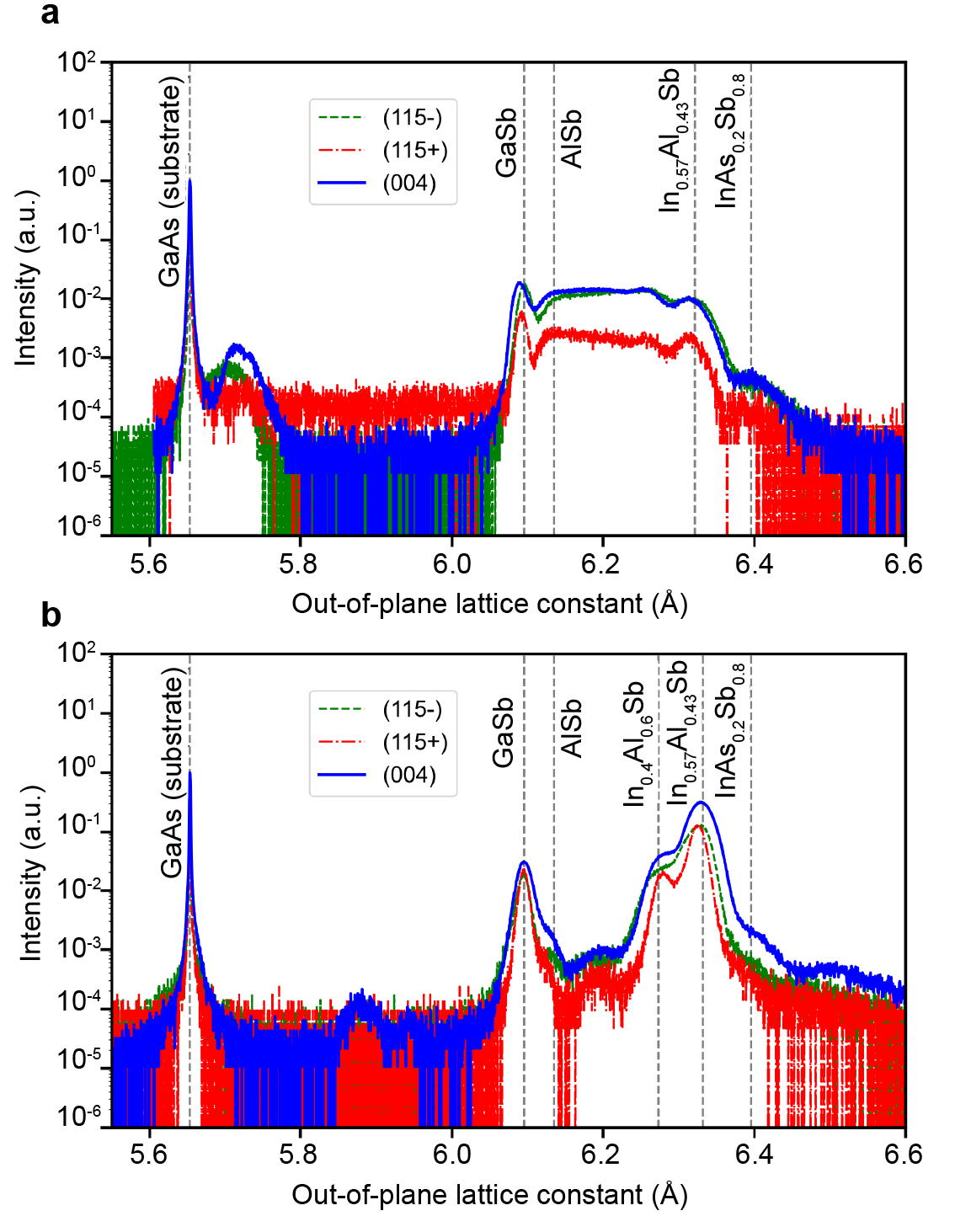}
    
    \caption{
    XRD measurement results for QWs with a graded buffer (a) and an InAlSb superlattice buffer (b), with the data converted to out-of-plane lattice constants.
}
\label{XRD}
\end{figure}

Figure \ref{XRD} compares the XRD measurements of InAsSb QWs with different buffer systems, measured along the (004), (115+), and (115–) directions. The $\omega -2\theta$ scans have been converted into out-of-plane lattice constants for clearer illustration and composition analysis. The active regions of both QWs were grown under identical conditions. From the lattice constants of InAsSb, we determine that both QWs have an As mole fraction of approximately $20\%$, and that the strain in InAsSb is negligible.

\section{Superconductor characterization}
\label{app:super}

\begin{figure}
    \centering
    \includegraphics[width = 1\linewidth]{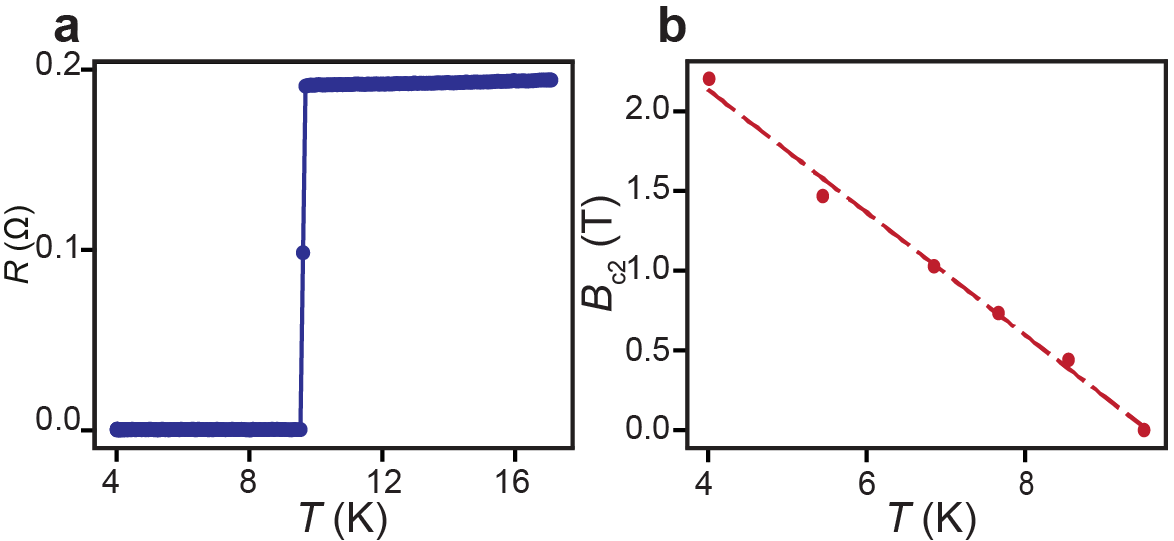}
    
    \caption{ Low-temperature  DC characteristics of the superconducting film. (a) $T_{\rm{c}}$  measurement. (b )$B_{\rm{c2}}$ as a function of temperature (b). The value of $B_{\rm{c2}}$ at $T=0$ K is extrapolated from the fit (dashed line). }
    \label{BcTc}
\end{figure}

Superconductor transport characterization is performed with In-contacted van der Pauw geometry using standard lock-in techniques (Fig. \ref{BcTc}). 
The critical temperature $T_{\rm{C}}$ is measured, and the out-of-plane critical magnetic field at 0 K $B_{\rm{c2}}(0)$ is determined to be larger than 3.5 T using the 
Werthamer–Helfand–Hohenberg relation~\cite{Werthammer1966}

\begin{equation}
    B_{c 2}(0)=\left.0.69 T_c \frac{d B_{c 2}}{d T}\right|_{T=T_c},
\end{equation}
which is presented by the dashed line.

\section{Effective mass measurement in a shallow InAsSb quantum well}

\begin{figure*}
\centering
\setlength\fboxsep{0pt}
\setlength\fboxrule{0pt}
\fbox{\includegraphics[width=6.5in]{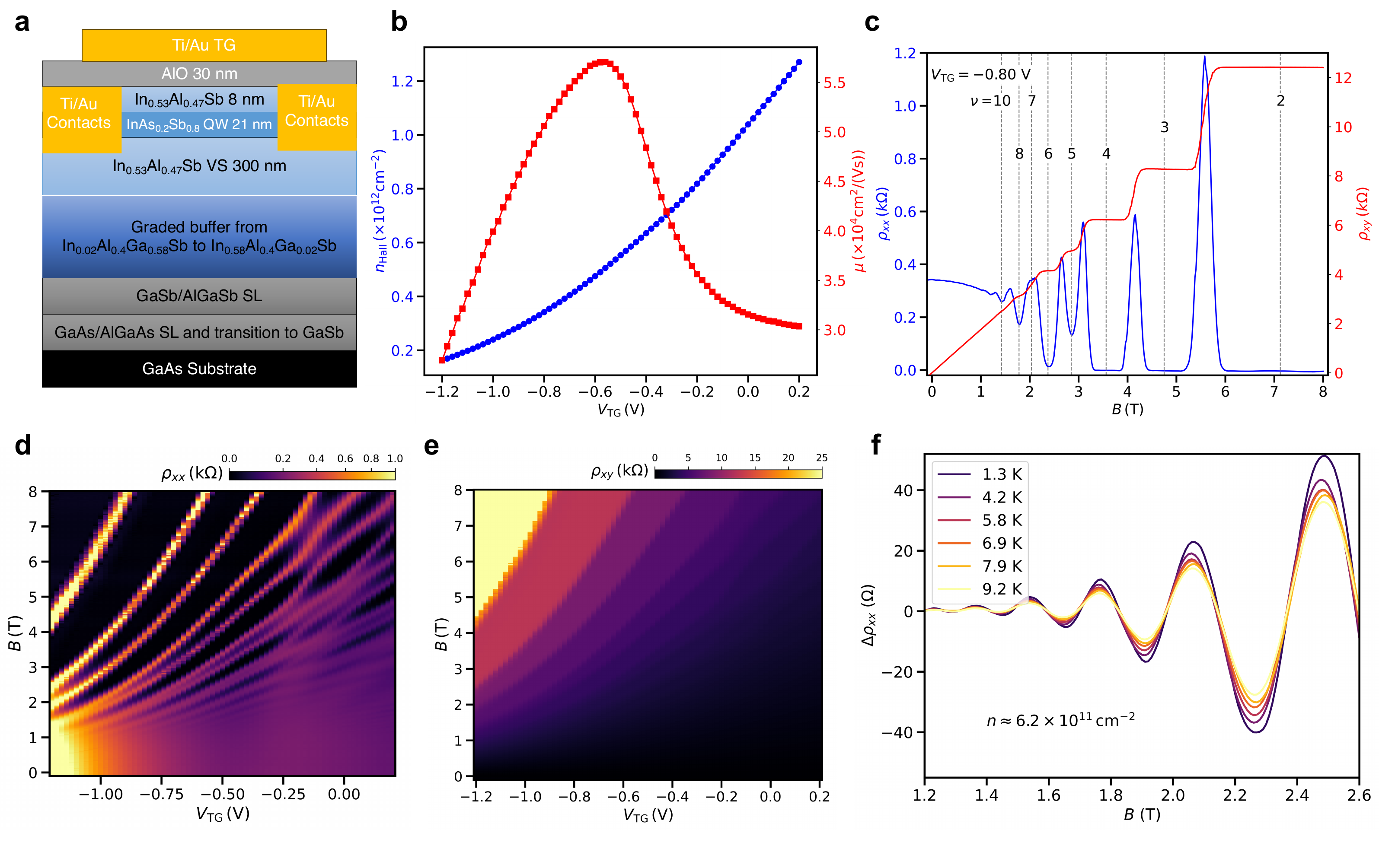}}
    \caption{
   (a) Schematic diagram of the shallow QW Hall bar device. The global top gate covers both the mesa and the contacts. (b) Hall density and mobility as functions of the top gate voltage, measured at a small magnetic field.
(c) SdH oscillations accompanied by the corresponding quantum Hall effect, measured at 15 mK with a top gate voltage of $-0.8\:\rm{V}$. The filling factors $\nu$ are labeled.
(d, e) Landau fan diagrams of the Hall bar, extracted from $\rho_{xx}$ and $\rho_{xy}$, respectively.
(f) Temperature dependence of the SdH oscillations at a Hall density of $6.2\times10^{11}\rm{cm^{-2}}$. 
The oscillatory component $\Delta$
$\rho_{xx}$ is obtained by subtracting a low-field polynomial background.
}

    \label{ShallowQW}
\end{figure*}
To measure the effective mass, we grew a shallow QW with the same growth conditions as the QW discussed in the main text, but with an 8 nm-thick In$_{0.53}$Al$_{0.47}$Sb capping layer deposited on top of the InAsSb layer. This capping layer significantly suppresses surface scattering, thereby enhancing the electron mobility.

As shown in Fig. \ref{ShallowQW}(a), the wafer was processed into a standard Hall bar with a global top gate covering both the mesa and the contacts. The sample fabrication process followed the procedure described in our previous publication \cite{Lei2023}. Fig. \ref{ShallowQW}(b) shows the Hall density $n$ and mobility $\mu$ as a function of top gate voltage $V_{\rm{TG}}$, determined from measurements near zero magnetic field. As $V_{\rm{TG}}$ increases, $n$ increases, while $\mu$ reaches a maximum value of 57,000 $\rm{cm^{2}/Vs}$ at $n = 5.2 \times 10^{11}\rm{cm^{-2}}$. The subsequent decrease in mobility at higher carrier densities is attributed to both interface scattering and the population of a second subband. In measurements over a wide magnetic-field range, the 2DEG exhibits standard Shubnikov–de Haas (SdH) oscillations accompanied by the corresponding quantum Hall effect. As shown in Fig. \ref{ShallowQW}(c), the SdH minima reach zero, while the Hall plateaus are quantized at the expected resistance values.

Figures \ref{ShallowQW}(d) and (e) show the longitudinal and transverse resistivities, $\rho_{xx}$ and $\rho_{xy}$, as functions of $V_{\rm{TG}}$ and magnetic field $B$. In both figures, the second fan diagram appears when $V_{\rm{TG}} > -0.3\:\rm{V}$ (Hall density exceeding $\sim 7 \times 10^{11}\:\rm{cm^{-2}}$), indicating the occupation of a second subband. In the regime where only one subband is occupied, the Hall density agrees well with the carrier density extracted from Shubnikov-de Haas (SdH) oscillations.

The effective mass of electrons in the InAsSb QW was determined from the temperature dependence of the SdH oscillations at $n = 6.2 \times 10^{11}\:\rm{cm^{-2}}$ (Fig. \ref{ShallowQW}(f)). By fitting the oscillation amplitude using the Ando formula, as in our previous work \cite{Bal2024,Lei2022,Lei2023}, the effective mass was extracted to be $m^* = 0.028\, m_e$, where $m_e$ is the electron mass in vacuum. Furthermore, we find that there is a noticeable band non-parabolicity of InAsSb within the measured density range, as a similar analysis at $n = 3.9 \times 10^{11}\rm{cm^{-2}}$ yields an effective mass of $m^* = 0.025\: m_e$.

\section{Weak antilocalization of InAsSb surface quantum well}
\label{App:semi}

\begin{figure*}
\centering
\setlength\fboxsep{0pt}
\setlength\fboxrule{0pt}
\fbox{\includegraphics[width=6.5in]{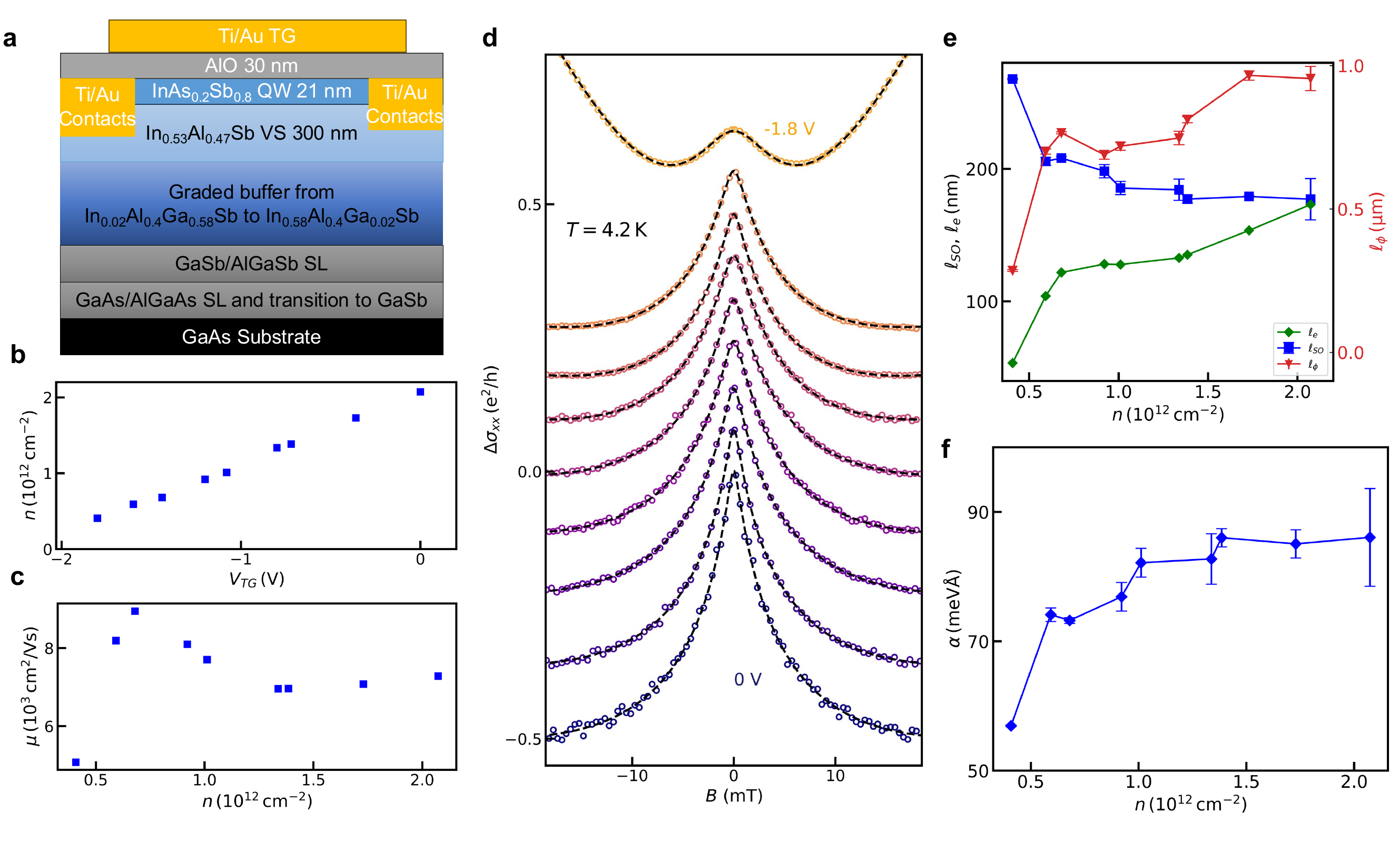}}
    \caption{
   (a) Schematic diagram of the surface QW Hall bar device. The global top gate covers both the mesa and the contacts. (b) Hall density as functions of the top gate voltage, measured at a small magnetic field. (c) Electron mobility as functions of the top gate voltage.
(d) WAL measured at different top gate voltages. A constant offset is applied to neighboring traces for a clear presentation. 
(e) Spin orbit length, phase coherent length and electron mean free path vs Hall density.
(f) Spin orbit coefficient vs Hall density.}
    \label{surfaceQW}
\end{figure*}

With the measured effective mass, the spin–orbit interaction (SOI) strength in the surface InAsSb QWs can be determined. Similar to the case of the shallow QW, the surface QW was characterized using a gated Hall bar device (Fig. \ref{surfaceQW}(a)).

Figure \ref{surfaceQW}(b) shows the Hall density $n$ as a function of top gate voltage $V_{\rm{TG}}$, which increases linearly with $V_{\rm{TG}}$. The mobility $\mu$ is plotted against $n$ in Fig. \ref{surfaceQW}(c), reaching a peak of $8.95 \times 10^3\: \rm{cm^{2}/Vs}$ at $n = 0.68 \times 10^{12}\rm{cm^{-2}}$, corresponding to a mean free path of $122\: \rm{nm}$. Based on the characterization of the shallow QW, we note that the second subband may be populated within this density range. However, due to the relatively high carrier density and low mobility, the transport properties of the two subbands are expected to be indistinguishable, since the $\rho_{xy}$ traces are always linear to $B$ within 2 T.

Weak antilocalization (WAL) measurements were performed at 4 K. The measurement and analysis methods follow those used in our previous publications \cite{Lei2022,Lei2023}. Figure \ref{surfaceQW}(c) shows the WAL correction to the longitudinal conductivity $\sigma_{xx}$ as a function of perpendicular magnetic field $B$. By fitting the data to the Hikami–Larkin–Nagaoka (HLN) model, we extract the spin–orbit length $l_{\rm{so}}$ and the phase coherence length $l_{\phi}$, and compare them to the mean free path $l_e$ in Fig. S3(d). In particular, $l_{\rm{so}}$ remains smaller than $l_e$ throughout the measurement range, indicating a strong spin–orbit interaction.

As $n$ increases, $l_{\phi}$ increases, while $l_{\rm{so}}$ decreases, suggesting that decoherence becomes weaker while SOI strength increases. To quantify the SOI strength independent of carrier density, the spin–orbit coupling coefficient $\alpha$ is calculated using the effective mass $m^* = 0.025m_e$ obtained from the shallow QW. As shown in Fig. \ref{surfaceQW}(e), $\alpha$ is large and increases with $n$ from 57 to 86 $\rm{meV\si{\angstrom}}$, highlighting the potential of this material system for realizing topological superconducting devices.

\section{Fabrication of hybrid devices}
\label{fab}
All the Josephson junctions (JJs) and SQUIDs presented in this work were fabricated on a single InAsSb–Nb chip. The device geometry was defined using electron beam lithography (EBL). After patterning the mesa, the Nb layer and the underlying InAsSb/InAlSb heterostructure were removed by reactive ion etching (RIE) and wet chemical etching, respectively. Nb was etched using a mixture of Ar and SF$_6$, while the InAsSb/InAlSb was etched in an aqueous solution of H$_2$O$_2$, H$_3$PO$_4$, and C$_6$H$_8$O$_7$. The semiconductor etching depth exceeded 100 nm to ensure the complete removal of any parallel conduction channels outside the mesa region.

Subsequently, the junction slits between the superconducting leads were defined by EBL, and Nb was selectively etched by RIE. The etching parameters, such as plasma power and gas dosage, were carefully optimized to fully remove the Nb layer without damaging the underlying InAsSb surface.

An AlO (3 nm) / HfO (20 nm) dielectric layer was deposited at low temperature sequentially, with AlO grown by thermal atomic layer deposition and HfO by plasma-enhanced atomic layer deposition. Finally, Ti/Au top gates were defined by electron beam evaporation and lift-off. All superconducting leads and metal gates were connected to the measurement setup using Al wire bonding.

\section{Numerical model for phase escape mechanisms in Josephson Junctions}

The phenomenological numerical model used in this work to describe the phase escape mechanism combines  descriptions of the macroscopic quantum tunneling (MQT) regime and the thermal activation (TA) regime. A sigmoid interpolation is used to describe the smooth crossover occurring between the MQT regime at low temperature and the TA regime at high temperature. The MQT regime is described as 
\begin{equation}
    \Gamma_{\text{MQT}} \propto \omega_p \exp\left( -\frac{7.2 \, \Delta U}{\hbar \omega_p}, \right)
\end{equation}

where $\Gamma_{\rm{MQT}}$, $\omega_p$, and $\Delta U$ are the phase escape rate in the MQT regime, plasma frequency, and potential barrier, respectively. Since the escape rate becomes nearly temperature-independent in this regime, this leads to a saturation of the switching current at low temperature with $I_{\text{sw}}(T \to 0) \to I_q$ where $I_{\rm{q}}$ is the quantum-limited switching current.

In the TA regime, the phase escape rate $\Gamma_{\rm{TA}}$ can be described with Kramer's law:
\begin{equation}
    \Gamma_{\text{TA}}(T) = a_{\text{TA}} \omega_p \exp\left( -\frac{\Delta U}{k_B T} \right),
\end{equation}
where $a_{\text{TA}}$ is a prefactor with a value of $\sim 1 / (2\pi)$, typically. 

For a tilted washboard potential with a cubic approximation near the saddle point, the barrier height scales as
\begin{equation}
\Delta U(I) \propto E_J \left( 1 - \frac{I}{I_c} \right)^{3/2},    
\end{equation}
where $E_J$ is the Josephson energy and $I_{\rm{c}}$ is the critical current.  
This results in the switching current $I_{\rm{SW}}$ that scales as
\begin{equation}
I_{\text{sw}}(T) \approx I_c \left( 1 - \left( \frac{T}{T_0} \right)^{2/3} \right)    
\end{equation}
where $T_0$ is a fitting parameter.

With the combination of MQT and TA regimes, the model can be described as

\begin{equation}
\begin{split}
& I_{\text{sw}}(T) =I_q + \\
  & \left\{ I_c \left( 1 - \left( \frac{T}{T_0} \right)^{2/3} \right) - I_q \right\} \cdot \left\{1 + \exp\left( -\frac{T - T^*}{\Delta T} \right)\right\}^{-1},
\end{split}
\end{equation}
with the crossover temperature $T^*$ and the width of the crossover region $\Delta T$ as fit parameters. 

\section{Gate tunability of the SQUID}
\begin{figure}
    \centering
    \includegraphics[width = 1\linewidth]{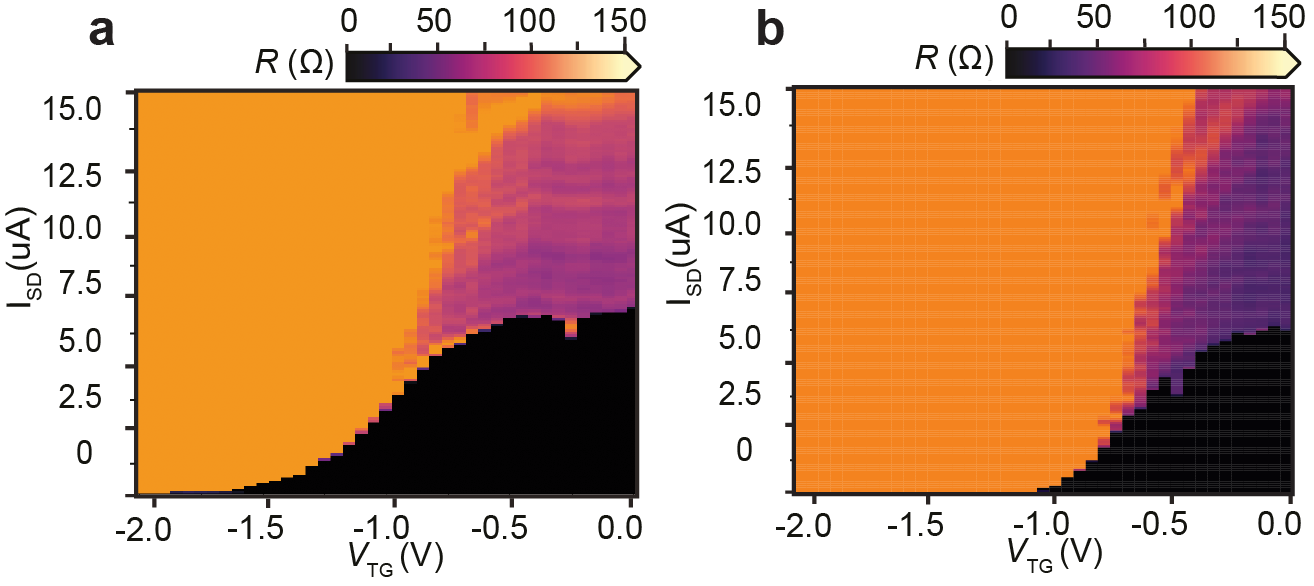}
    \caption{
    The resistance $R$ as a function of bias current $I_{\rm{SD}}$ and top gate voltage $V_{\rm{TG}}$ for the small junction (a) and the large junction (b) on the SQUID.}
\label{sup_squid}
\end{figure}

Figure \ref{sup_squid} shows the resistance $R$ as a function of the top gate voltage $V_{\rm{TG}}$ and the bias current $I_{\rm{SD}}$ for both junctions on the SQUID discussed in the main text. A voltage of $V_{\rm{TG}}=-2\:  \rm{V}$ is sufficient to fully pinch off both junctions. 
\label{app:squid}

\section{Additional devices}
\label{App:otherJJ}
\begin{figure}
\centering
\includegraphics[width = 1\linewidth]{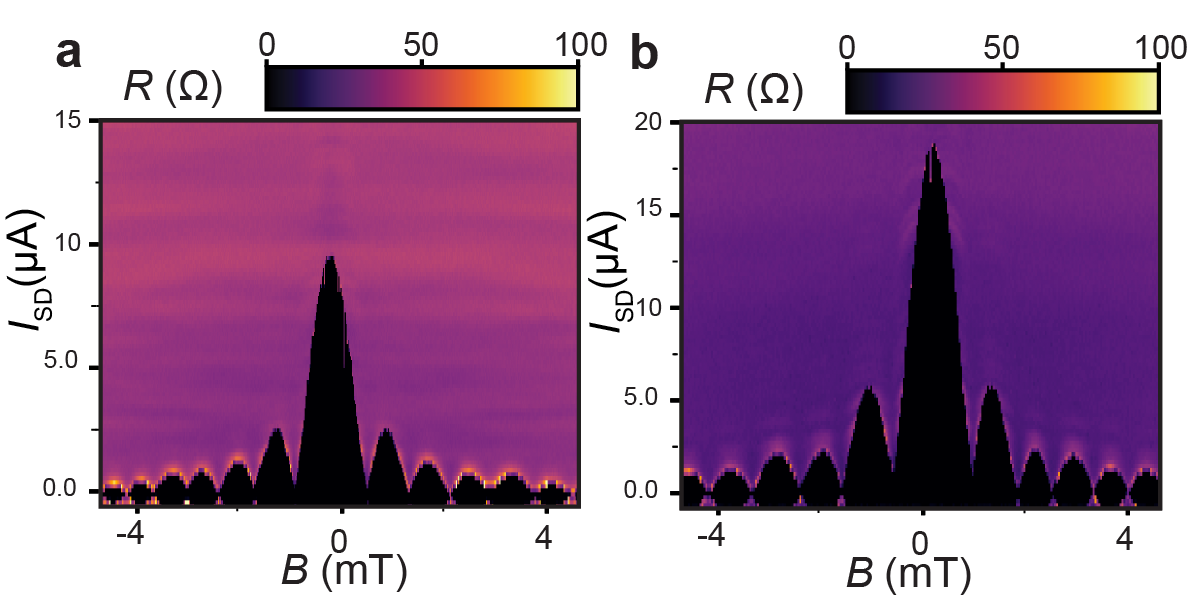}
\caption{
(a) The measured resistance $R$ as a function of bias current and magnetic field for a JJ of size 100 nm by 4 $\mu$m (a) and 75 nm by 4 $\mu$m (b).}
\label{sup:moreJJ}
\end{figure}

Figure \ref{sup:moreJJ} shows the Fraunhofer diffraction patterns of two additional JJs with sizes of 100 nm by 4 $\mu$m (a) and 75 nm by 4 $\mu$m (b). The diffraction patterns are qualitatively similar to the ones discussed in the main text and further substantiate the high critical current densities achieved in these devices. Due to the different sizes in geometry, these two JJs are different in the critical current and spacing of the nodes in magnetic fields.

\end{document}